\documentclass[journal]{IEEEtran}
\IEEEoverridecommandlockouts
\newcommand{\subparagraph}{}
\usepackage[american,fulldiode]{circuitikz}
\usepackage{epstopdf}
\usepackage{caption}
\usepackage{steinmetz}
\usepackage{gensymb}

\usepackage[cmex10]{amsmath}
\usepackage{array}
\usepackage{mdwmath}
\usepackage{mathrsfs}
\usepackage{mdwtab}
\usepackage{caption}
\usepackage[caption=false,font=footnotesize]{subfig}
\usepackage{cite}
\usepackage{amssymb}

\newcommand{\R}{\mathbb{R}}

\usepackage{blindtext, subfig}
\usepackage{multicol}
\usepackage{amssymb}

\mathcode"65"65
\def\bad{\spaceskip=0.33emplus0.6emminus0.15em\immediate\write5{\string\bad}}

\def\({\left(}
\def\){\right)}

\def\bad{\spaceskip=0.33emplus0.6emminus0.15em\immediate\write5{\string\bad}}

\def\bad{\spaceskip=0.33emplus0.6emminus0.15em\immediate\write5{\string\bad}}

\def\({\left(}
\def\){\right)}

\def\bad{\spaceskip=0.33emplus0.6emminus0.15em\immediate\write5{\string\bad}}

\begin{document}

\title{Embedding AC Power Flow in the Complex Plane Part II: A Reliable Framework for Voltage Collapse Analysis}

\author{Sina S. Baghsorkhi, \textit{Member, IEEE}, and Sergey P. Suetin\vspace{-1mm}}

\maketitle
\pagenumbering{gobble}
\begin{abstract}
Part II of this paper elaborates on the unique capability of the proposed power flow analysis framework to obtain the true solution corresponding to the stable operating point of a network. It explains the significance of obtaining the true solution for an accurate assessment of the voltage collapse margin. This feature distinguishes the framework from all iterative and non-iterative heuristic approaches as demonstrated in the context of a 7-bus network with Newton-Raphson, its variants and semidefinite and moment-based relaxations of power flow.
Another important feature of this framework is that it obtains the true solution when it exists and declares its non-existence otherwise. This is demonstrated in the context of small networks and in comparison with heuristic approaches. This paper also explores how the proposed framework detects a limit-induced bifurcation where a network controller reaching its limit can initiate voltage collapse.
\vspace{0mm}
\end{abstract}

\begin{IEEEkeywords}
AC power flow, voltage collapse, saddle-node bifurcation, limit-induced bifurcation. \vspace{-1mm}
\end{IEEEkeywords}

\section{Introduction}

The study of voltage instability and developing proximity metrics to saddle-node bifurcation\footnote{This term was first used in the context of power system equations as ``fold bifurcation" in reference~\cite{Sastry}.} in power system was pioneered in the USSR by V.A. Venikov and his colleagues~\cite{Venikov0,Venikov1,Venikov2,Venikov} which built on the earlier works of P.S. Zhdanov and O.G.C. Dahl on the topic of power system stability in the 1930s-40s~\cite{Zhdanov0, Zhdanov1,Dahl}. The significance of these topics in relation to the mathematical modeling of AC power flow were not fully understood outside the USSR until 1980s-1990s when numerous papers were published in the US and Japan on these subjects. From a mathematical point of view the topic of voltage collapse is intimately connected with the solutions of the AC power flow problem and this had been noted by Venikov~\cite{Venikov} and others~\cite{4074830,Iwamoto,Klos}.

Voltage collapse is a dynamic phenomenon the study of which involves time-domain analysis of the full dynamical model of power systems~\cite{Chiang}. Hence we need to clarify the relationship between voltage collapse and the singularity analysis of the Jacobian of power flow algebraic equations pioneered by Venikov and adopted unanimously in power system engineering afterwards. Consider the following generic dynamical system,\vspace{-3mm}

\begin{align}
\dot{x}&=f(x,\lambda)\label{DS}
\end{align}

The equilibria of this system are given as the solutions of $0=f(x,\lambda)$ and as $\lambda$, the set of parameters, varies over time the structure of the solutions may undergo a qualitative change or bifurcation, such as the merging of two solution branches, one stable (in a local sense) and the other unstable and their subsequent disappearance. As the system reaches bifurcation the equilibrium point may no longer be asymptotically (or locally) stable. Note that the algebraic equations are sufficient to detect the onset of instability but the dynamic evolution of states along the unstable manifold of the equilibrium involves time-domain analysis.

Power system operation often requires a much more complicated dynamical model than~\eqref{DS}. One such detailed model is given by a system of autonomous nonlinear differential-algebraic equations (DAEs):\vspace{-3mm}

\begin{subequations}\label{DAE}
\begin{align}
\dot{x}&=f(x,y,z,\lambda)\label{diff}\\
0&=g(x,y,z,\lambda)\label{algeb}\\
0&=h(z,\lambda)\label{lf}
\end{align}
\end{subequations}

\noindent
where $x$ and $y$ are the dynamical and algebraic states associated with electromechanical devices in the network such as synchronous and doubly-fed induction generators and their (AVR, rotor speed, pitch-angle) controllers, $z$ the set of power flow variables and $\lambda$ the set of parameters. As demonstrated by Sauer and Pai~\cite{Sauer0}, the standard set of power flow equations or its variants represented in~\eqref{lf}, can always be solved for the {\em multi-valued} $z_0$ independent of initial conditions of dynamical states and other algebraic variables. Once a particular  $z_0$ is obtained, \eqref{diff}-\eqref{algeb} can be solved for the corresponding equilibrium point $E_0=(x_0,y_0,z_0)$. From implicit function theorem it follows that when the Jacobians $g_y(x,y,z,\lambda)$ and $h_z(z,\lambda)$ are nonsingular there exists a smooth function $F$ such that $\dot{x}=F(x,\lambda)$ similar to~\eqref{DAE}. Note that algebraic variables are eliminated in $F$. Under certain assumptions\footnote{These assumptions remove the possibility of oscillatory instability or Hopf bifurcation.} on the dynamical models, the differential-algebraic system of~\eqref{DAE} the nonlinear dynamics of which is represented locally by $F$ can experience bifurcation if, and only if, $h_z$ is singular~\cite{Venikov,Sauer0}. Note that the singularity of the power flow Jacobian $h_z$ implies that the obtained solution of $z_0$ is on the solution space boundary. From a geometric point of view this is a branch point in the parameter space of~\eqref{lf} where at least two algebraic sheets coalesce. As the power flow parameters are perturbed there is a structural change in the set of equilibria containing $E_0$ where at least two equilibria coalesce into a single equilibrium and disappear. In other words, at a bifurcation (branch) point ``two solutions are born or two solutions annihilate each other"~\cite{Seydel}. More precisely at the saddle-node bifurcation point the following conditions hold between $z_0 \in \R^n$ and a given bifurcation parameter value $\alpha_0 \in \R$ where $\alpha \subset \lambda$~\cite{Seydel}:\vspace{2mm}

\begin{description}
  \item[(1)] $h(z_0,\alpha_0)=0$
  \item[(2)] $\text{rank}$ $h_z(z_0,\alpha_0)=n-1$
  \item[(3)] $h_{\alpha}(z_0,\alpha_0) \notin$ $\text{range}$ $h_z(z_0,\alpha_0)$, i.e. $\text{rank}$ $(h_{\alpha}(z_0,\alpha_0) | h_z(z_0,\alpha_0))=n$
  \item[(4)] there exists a parametrization $z(\sigma)$, $\alpha(\sigma)$ with $z(\sigma_0)=z_0$, $\alpha(\sigma_0)=\alpha_0$ where $\text{d}^2\alpha(\sigma_0)/\text{d}{\sigma}^2\neq0$ \end{description}

\noindent
Note that condition (2) is the singularity of the power flow Jacobian. Condition (3) ensures that the bifurcation point is not simply an intersection of two branches but truly a turning point, i.e. the sensitivities of elements of $z$ to $\alpha$ is infinite. Condition (4) is to rule out the degeneracy of the saddle-node bifurcation~\cite{Seydel}. At saddle-node bifurcation the asymptotical stability of the equilibrium point is lost and this  signals the {\em onset} of voltage collapse phenomenon. However the dynamic evolution of voltage collapse involves time-domain analysis of the full DAEs. What even further complicates the ensuing analysis is the presence of limits of controllers and protective schemes~\cite{Mitigation} which changes the structure of DAEs as voltages dynamically collapse. So the fact that the system experiences saddle-node bifurcation may not necessarily entail a catastrophic outcome such as a black-out. Therefore the aim of this paper is to detect the {\em onset} of voltage collapse phenomenon as pioneered by Venikov and his colleagues~\cite{Venikov}. In this paper by operating point we refer to $E_0=(x_0,y_0,z_0)$ which is uniquely characterized by $z_0$. By a stable operating point we mean an equilibrium point $E_0$ that is asymptotically stable and is adequately represented by $z_0$ the true solution of the AC power flow, if it exists.

With the above explanations it should be clear that the AC power flow problem or $0=h(z,\lambda)$, characterized by a nonlinear system of equations that describe the steady-state operation of a network, is the most fundamental problem in power system engineering in the sense that its correct analysis is vital for the stable operation of power systems. This problem has many solutions. Most of these solutions are false and can not be physically realized. There are some solutions that can be realized but correspond to (locally) unstable equilibria of the system of~\eqref{DAE}.  Obtaining and characterizing these solutions involves numerical analysis of the flow of electric power in an interconnected system. The existence of multiple solutions creates challenges for voltage stability studies and developing a proximity index to the onset of voltage collapse process. If we can correctly solve the power flow equations and determine the true operating point of a network then we can glean some information on the stability margin of that operating point and prepare to implement the right set of control measures to increase that margin and prevent voltage instability or collapse.

The power flow problem is currently solved by the classical Newton's method or its variants. This
involves successive linearization of the equations and approximation of the solution starting from an
initial guess. If the solution obtained at each iteration converges and the mismatch error of the equations
is lower than a certain tolerance, the approximated solution is declared as the solution of the
power flow problem. The Jacobian matrix which is formed by linearizing the equations at this approximated
solution contains information that can be further processed as proximity indices to voltage collapse.
For example as the operating point moves toward the bifurcation boundary, the condition
number of the Jacobian rapidly increases which means that the smallest eigenvalue of the Jacobian
tends toward zero. So the smallest eigenvalue can be a \emph{very crude} indicator of how close a given operating
point is to the onset of voltage collapse. A similar index can be developed based on the singular values of the
Jacobian matrix. Unfortunately Newton's method, as robust as it is, may still fail to converge or it may converge to undesirable solutions. There is a general consensus among power flow experts that the existing numerical methods are likely to exhibit anomalies when the power system is under stress.

Commercial developers of power flow software often claim that their software can reliably determine the correct operating point provided that it is feasible. However these claims are not supported by mathematics~\cite{Newton} and are even contradicted by numerous papers published in 1960s through 1990s~\cite{4074830,4110791,Iwamoto,4181598,Klos,Johnson}. Reference~\cite{Johnson} is among the earliest studies that show some of these false or unstable solutions are virtually indistinguishable from the stable operating point, i.e. the voltage magnitudes of these undesirable solutions seem reasonably high and normal. Numerical algorithms used in power flow studies, including Newton's method itself, are all based on heuristics. Therefore when these algorithms do not converge no conclusion can be made on whether the power flow problem is truly infeasible or these algorithms have failed to obtain the solution. In this regard, an extremely valuable document was produced by the developers of the most widely used commercial power flow software, PSS/E, at Siemens~\cite{PSSE}. What is particularly significant in this document is the reference to FACTS devices and other network controllers such as tap-changers and phase-shifters as other potential sources of numerical issues. The algebraic constraints introduced by these controllers make the convergence of numerical algorithms even more problematic than in the case of stressed power systems. Most modern power electronic devices ``introduce highly nonlinear equations" which should be suitably initialized to ensure convergence when using the Newton's method~\cite{FACTS,FACTS2}. Unfortunately there does not exist a methodology of initialization that would guarantee convergence to a physically meaningful solution or convergence at all (see for example reference~\cite{FACTS} for the difficulties related to the initialization of these devices in Newton's power flow method). Thus with the emergence and more frequent usage of these controllers in modern power systems the shortcomings of traditional numerical frameworks are much more noticeable. The transformation of modern power systems with highly variable generation and a new category of network controllers with sophisticated control capabilities requires a reliable numerical framework to determine the state of the network and its stability margin. This new numerical framework should overcome the shortcomings of the existing frameworks and be competitive in terms of computational performance.

The Part II of this paper is organized as follows. In Section II we demonstrate the significance of obtaining the true operating pointing of a network for voltage collapse analysis. This is shown in the context of a 6-bus network where the inspection of voltage magnitudes and angles tend to pick the false solution of the power flow over the true solution as the normal operating point of the network. We also show how the two different embedding approaches, discussed in Part I of this paper, filter out false solutions whenever no stable solution exists. In Section III we demonstrate the superiority of the proposed framework of embedding the AC power flow over conventional and recently developed methods of solving the power flow problem.  We introduce a 7-bus network where Newton-Raphson either fails to converge or converges to false solutions as the active power output of a given generator changes. We also show that in this network, first-order semidefinite relaxation only obtains the solution in a small subset of the stable solution branch and that the second-order (moment-based) relaxation obtains the false solution branches. The numerical results of these methods are contrasted with that of the embedding framework which consistently obtains the stable solution whenever it exists and declares the non-existence of a physically meaningful solution beyond the saddle-node bifurcation. In this network, the zero-pole distribution of the Pad\'{e} approximants confirms the general pattern of voltage stability margin observed in Part I of this paper. We also address a deep-rooted confusion on the prospect of a combination of heuristic approaches finding the true solution. We specifically analyze two modifications of the Newton-Raphson that, to a large part, address the shortcomings of conventional Newton-Raphson with a flat start in the context of the 7-bus network but fail when the network is slightly modified. In Section IV we explain how the embedding framework detects voltage collapse instigated by a reactive device reaching its limit or limit-induced bifurcation and demonstrate that in the context of previously introduced 7-bus network.


\begin{figure*}
\begin{circuitikz}[scale=1, transform shape]

\path[draw,line width=4pt] (0,0) -- (0,2);
\draw (-1.6,2.2) node[right] {\text{Reference}};
\draw (-1.6,1.8) node[right] {\text{(Slack)}};
\draw (-1.6,-0.3) node[right] {$|V_r| = 1.00$};
\draw (-1.6,0.2) node[right] {$\theta_r = 0^\circ$};
\path[draw,line width=1pt] (-0.5,1) -- (0,1);
\draw[line width=1] (-1,1) circle (0.5);
\draw (-1,1) node{\scalebox{2}{$\sim$}};

\path[draw,line width=4pt] (4,0) -- (4,2);
\draw (3.8,2.25) node[right] {1};
\draw[-latex,line width=1pt] (4,1)-- (5,1);
\draw (4.90,0.98) node[right] {$0.25+j0.10$};

\path[draw,line width=4pt] (9,0) -- (9,2);
\draw (8.8,2.25) node[right] {2};
\draw[-latex,line width=1pt] (9,1)-- (10,1);
\draw (9.9,0.98) node[right] {$0.35+j0.10$};

\path[draw,line width=4pt] (14,0) -- (14,2);
\draw (13.8,2.25) node[right] {5};
\path[draw,line width=1pt] (14.5,1) -- (14,1);
\draw[line width=1] (15,1) circle (0.5);
\draw (15,1) node{\scalebox{2}{$\sim$}};
\draw (14,0.25) node[right] {$|V_5| = 1.10$};
\draw (14,1.75) node[right] {$P_5 = 1.00$};


\path[draw,line width=4pt] (9,-4) -- (9,-2);
\draw (8.8,-1.75) node[right] {4};
\path[draw,line width=1pt] (9,-3) -- (9.5,-3);
\draw[line width=1] (10,-3) circle (0.5);
\draw (10,-3) node{\scalebox{2}{$\sim$}};
\draw (9,-3.75) node[right] {$|V_4| = 1.10$};
\draw (9,-2.25) node[right] {$P_4 = 0.90$};

\path[draw,line width=4pt] (14,-4) -- (14,-2);
\draw (13.8,-1.75) node[right] {3};
\draw[-latex,line width=1pt] (14,-3)-- (15,-3);
\draw (14.9,-3.02) node[right] {$0.35+j0.20$};
%

\path[draw,line width=1pt] (0,1)-- (4,1);
\draw (2,1.55) node[below] {$0.70+j0.40$};

\path[draw,line width=1pt] (4,1.66)-- (9,1.66);
\draw (6.5,2.20) node[below] {$0.50+j0.50$};

\path[draw,line width=1pt] (9,1.66)-- (14,1.66);
\draw (11.5,2.20) node[below] {$0.40+j0.50$};


\path[draw,line width=1pt] (9,0.33) -- (8.5,0.33);
\path[draw,line width=1pt] (8.5,0.35) -- (8.5,-2.35);
\path[draw,line width=1pt] (8.5,-2.33) -- (9,-2.33);
\draw (8.25,-1) node[rotate=90] {$0.30+j0.50$};


\path[draw,line width=1pt] (14,0.33) -- (13.5,0.33);
\path[draw,line width=1pt] (13.5,0.35) -- (13.5,-2.35);
\path[draw,line width=1pt] (13.5,-2.33) -- (14,-2.33);
\draw (13.25,-1) node[rotate=90] {$0.60+j0.80$};

\path[draw,line width=1pt] (9,0.334) -- (9.51,0.334);
\path[draw,line width=1pt] (9.5,0.334) -- (13.5,-3.667);
\path[draw,line width=1pt] (13.49,-3.667) -- (14,-3.667);
\draw (11.5,-1.25) node[rotate=-45] {$0.50+j0.80$};

\end{circuitikz}
\caption{6-bus network\vspace{7mm}}\label{6Bus}

 \centering
        \includegraphics [scale=1.1,trim=0.0cm 0.cm 0.0cm 0.0cm,clip]{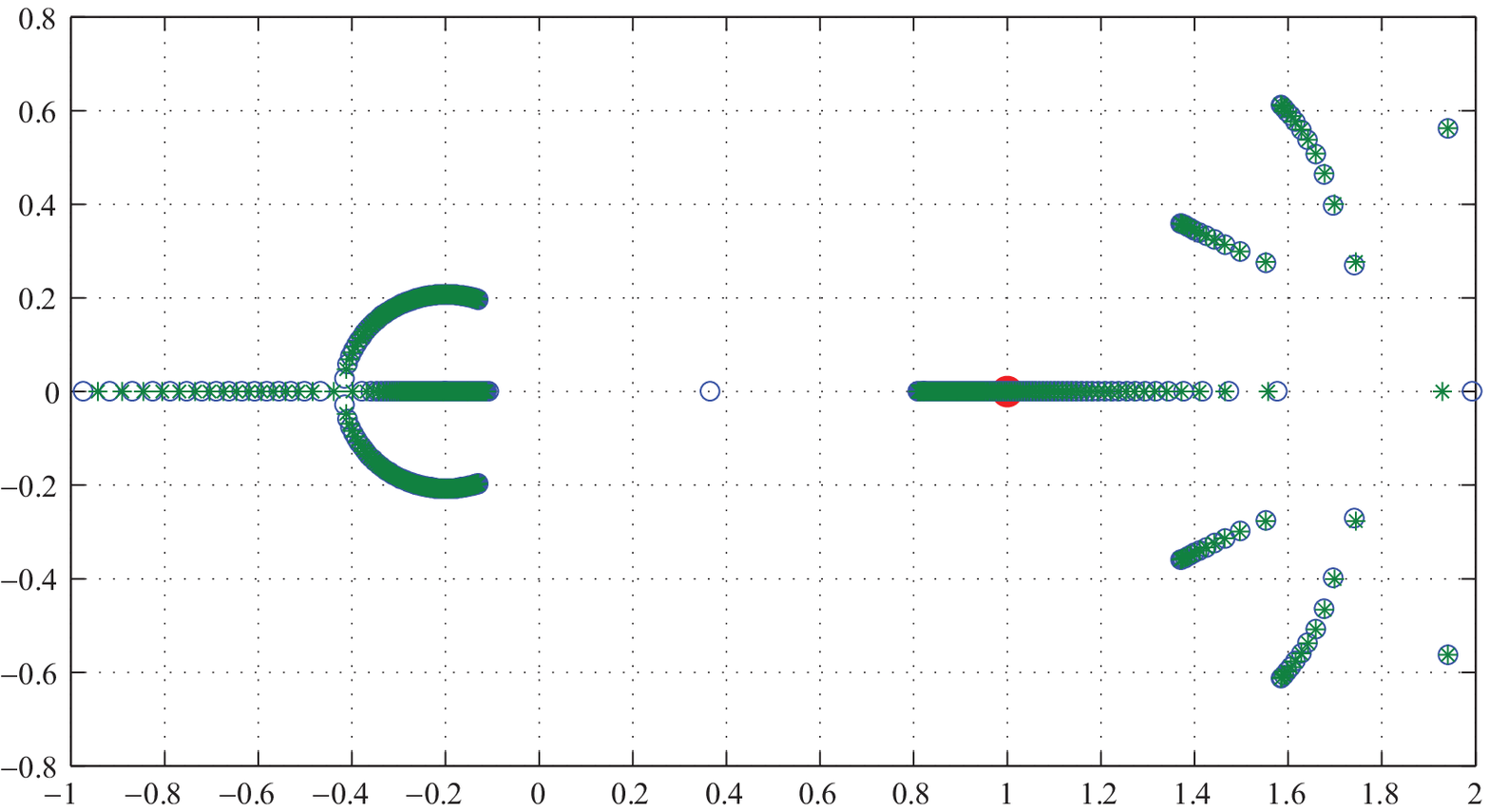}
         \caption{Zero-pole distribution of PA[1000/1000] depicting the analytic structure of voltage phasors in the network of Fig.~\ref{6Bus} (corresponding to
the embedding approach defined in Section III of Part I of this paper).\vspace{5mm}}\label{PadeTF}

  \centering
        \includegraphics [scale=1.1,trim=0.0cm 0.cm 0.0cm 0.0cm,clip]{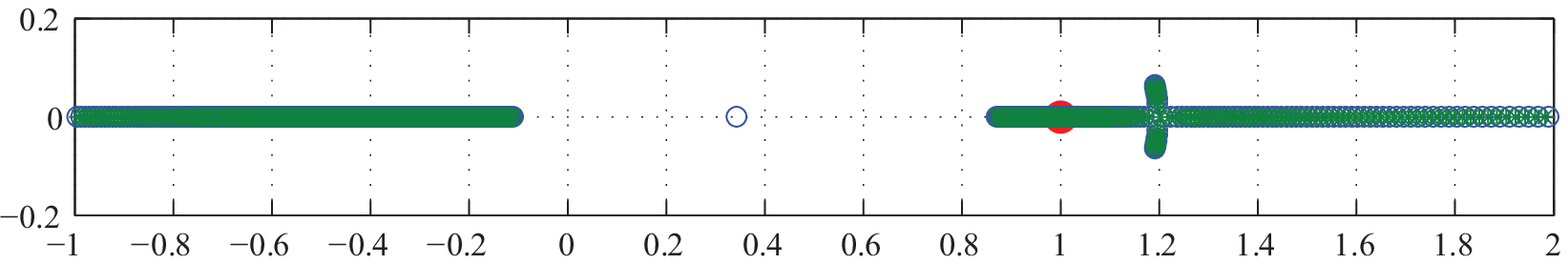}
         \caption{Zero-pole distribution of PA[1000/1000] depicting the analytic structure of voltage phasors in the network of Fig.~\ref{6Bus} (corresponding to
the embedding approach defined in Section IV of Part I of this paper).\vspace{0mm}}\label{PadeTF_2nd}

\end{figure*}

\section{Distinction between True (Stable) and False Solutions in the context of hybrid dynamical systems}

The first step to restore a network from a blackout or a planned outage is to energize it via a strong voltage source. This voltage source is either a black-start generator or the tie-lines of a neighboring external network. Only after the transmission lines are energized is it possible to pick up load and add generation. From this state of no-load and no-generation the network can be stressed by increasing the load and generation in any direction\footnote{In practice this direction can not be arbitrary but rather is constrained by the topology of the network and the distribution of load and generation.}  while maintaining a stable operating point until some type of bifurcation and voltage instability occurs. Hence the system can be moved from the initial unstressed state to any normal stable operating point, corresponding to a solution of the power flow equations without encountering limit-induced or saddle node bifurcations. Even as the topology of the network changes, by line outages, the preservation of power system stability indicates that there is path that connects the unstressed state of the modified network to the new operating point.

Despite the fact that its existence is guaranteed such a path in the power flow parameter space, comprised of load and generation, that connects a given stable operating point to an unstressed energized state may be a highly non-trivial one. {\bf Nonetheless a true, i.e. stable, operating point of a network is uniquely\footnote{ The uniqueness is only true in the absence of discrete events. In practice as a network is stressed from the initial state to a final point in the power flow parameter space there are often some controllers that reach their limits. The exact set and the order in which these limits are encountered depends on the specific path taken in the parameter space.} and unambiguously characterized in relation to this unstressed energized state and thus distinguished from all other operating points that can only be realized, if at all, in a transient state.}

We should emphasize that a true or stable solution is not necessarily unique due to power systems being essentially hybrid dynamical systems characterized by both continuous and discrete states and above all by discrete events or triggers~\cite{Hiskens}. These discrete events includes changing of control modes in generators, static VAR controllers, tap changing transformers and other FACTS devices. Each configuration of these discrete events or triggers produces a different set of power flow equations, so that there are multiple possible stable operating points for a given combination of load and generation. Thus the claim in reference~\cite{Trias} that ``load flow equations have multiple solutions,
and only one of them corresponds to the real operative state of the electrical system" is inaccurate.

\begin{figure*}
        \centering
\includegraphics[scale=0.69,trim=2.0cm 0cm 0.0cm 1.0cm,clip]{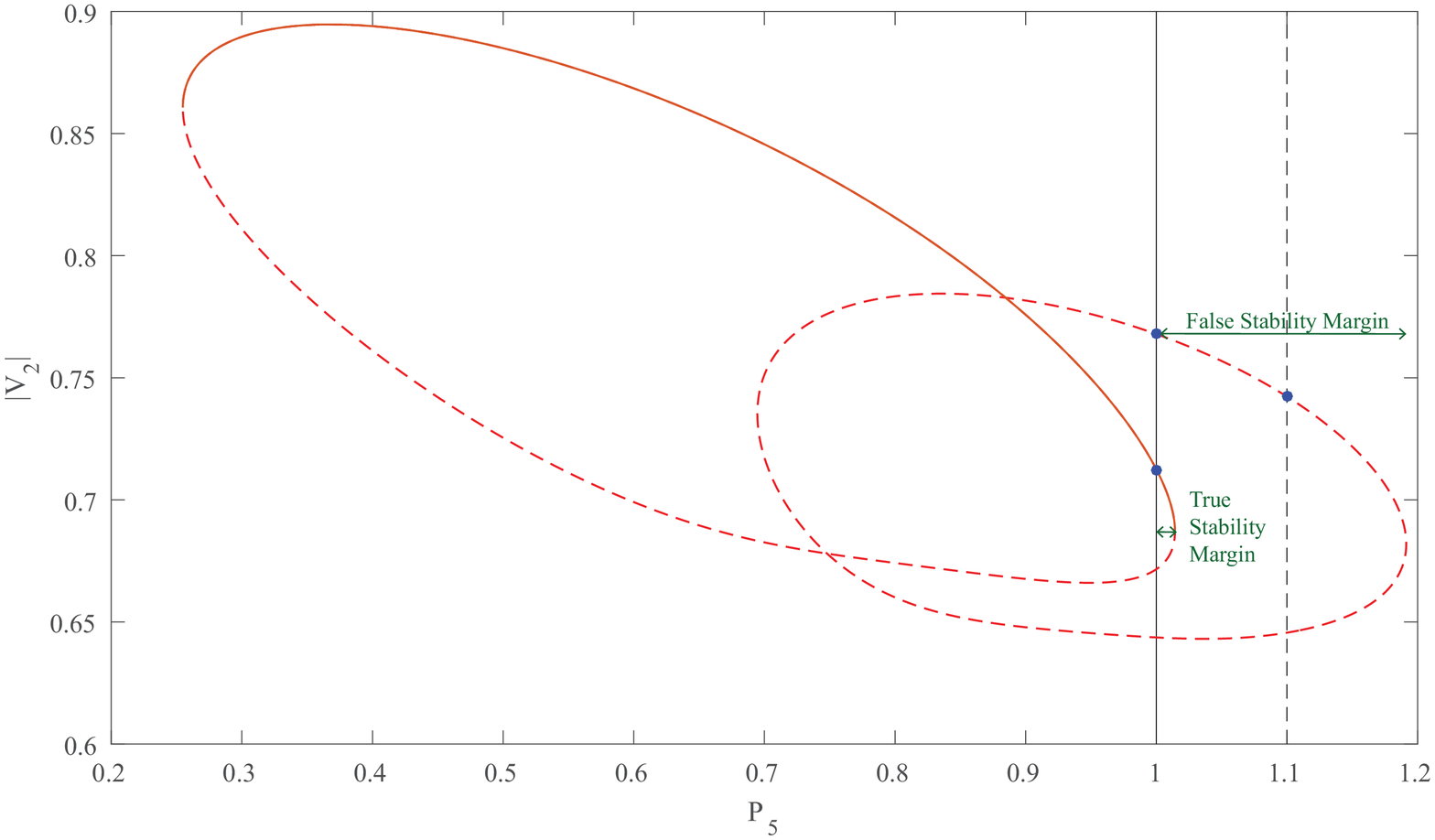}
\caption{Stable (solid) and unstable (dotted) solution branches of the network of Fig.~\ref{6Bus}\vspace{-2mm}}\label{TFB}
\end{figure*}


\begin{table}[h]
\caption{True versus False Solutions}\label{TFT}
\centering
\renewcommand{\arraystretch}{1.25}

\begin{tabular}{|c|l|l||c|l|} \hline
Voltage   &  \multicolumn{2}{c||}{\hspace{-1mm}Operating Point ($P_5=1.00$)}      & \multicolumn{2}{c|}{\hspace{-1mm}Operating Point ($P_5=1.10$)}  \\ \cline{2-5}
Phasor    &\multicolumn{1}{c|}{True}  & \multicolumn{1}{c||}{False}   & \multicolumn{1}{c|}{True}  & \multicolumn{1}{c|}{False}  \\\hline
$V_1$& ${\bf 0.42}$\phase{37\degree} & ${\bf 0.72}$\phase{19\degree} & $\hspace{6mm}-\hspace{6mm}$ & ${\bf0.65}$\phase{26\degree}\\\hline
$V_2$& ${\bf 0.71}$\phase{95\degree} & ${\bf 0.77}$\phase{45\degree} & $\hspace{6mm}-\hspace{6mm}$ & ${\bf0.74}$\phase{61\degree}\\\hline
$V_3$& ${\bf 0.58}$\phase{90\degree} & ${\bf 0.30}$\phase{28\degree} & $\hspace{6mm}-\hspace{6mm}$& ${\bf0.32}$\phase{47\degree}\\\hline
$V_4$& $1.10$\phase{119\degree}      & $1.10$\phase{69\degree} &       $\hspace{6mm}-\hspace{6mm}$ & $1.10$\phase{85\degree}\\\hline
$V_5$& $1.10$\phase{102\degree}      & $1.10$\phase{46\degree} &       $\hspace{6mm}-\hspace{6mm}$& $1.10$\phase{66\degree}\\\hline
$V_{\text{r}}$&$1.00$\phase{0\degree} & $1.00$\phase{0\degree} &       $\hspace{6mm}-\hspace{6mm}$ & $1.00$\phase{0\degree}\\\hline
\end{tabular}
\end{table}

Table~\ref{TFT} contrasts the solution corresponding to the true stable operating point of the network of Figure~\ref{6Bus} with a false solution. This is a 6-bus network with 3 load (PQ) buses labeled 1-3, two generator (PV) buses labeled 4 and 5 and a reference (slack) bus. All values are in per unit. Line and load parameters are indicated as complex quantities. The generator voltage magnitudes are controlled at 1.10 and the active power output at bus 4 and bus 5 are 0.90 and 1.00. Notice that between the two solutions the false solution has more of the hallmarks of a normal operating point. The voltage magnitudes are higher and phase-angle differences between adjacent buses are contained in a smaller range. The distinct solution branches are demonstrated in Figure~\ref{TFB} when active power generated at bus 5 is set free. The embedding framework discovers only the solid branch. The solutions on dotted branches\footnote{We will demonstrate in a future publication that these operating points are asymptotically unstable by applying time-domain analysis to the detailed dynamical model of the system.} are unstable. What is significant in this analysis is the {\em non-existence} of a stable operating point when active power generation at bus 5 is increased to 1.10. Some commercial power flow software return a false solution for this set of power flow parameters, especially when the power flow is initialized not from flat-start but from a previously-solved stable state with parameters lying in the vicinity of those considered, i.e. $P_5=1.10$. This is a shortcoming of existing power flow packages. In contrast, the embedding framework declares the non-existence of a true solution when $P_5=1.10$. This is clearly demonstrated by the zero-pole distribution of Pad\'{e} approximants, for the two different embedding approaches defined in Part I of this paper, in Figures~\ref{PadeTF} and~\ref{PadeTF_2nd} where the analytic arcs on the real axis have completely covered $z=1$. This means that there is absolutely no way to increase the loading and generation in the system from an energized unstressed state to the new condition corresponding to $P_5=1.10$ without first encountering some voltage instability. The power flow software packages that return such solutions provide the network operator with a false sense of security because the given operating point (see Figure~\ref{TFB}) appears to enjoy a wide voltage collapse margin at $P_5=1.10$ but as the generation at Bus 5 increases from 1.00 toward 1.10, at some point, the network undergoes a sudden voltage transition.

\begin{table}
\caption{$P_6=0.20$ (all methods finding the stable solution)} \label{t1}
\vspace{-1mm}
\centering
\renewcommand{\arraystretch}{1.2}
\begin{tabular} {|c||c|c|c|c|} \hline 
 Voltage        & Pad\'{e}      & Newton & SDP  & SDP \\
 Magnitude      & Approx.      & Raphson & (1st order)  & (2nd order) \\\hline
$|V_1|$&0.9408  &0.9408 &0.9408 &0.9408 	\\\hline
$|V_2|$&0.9774  &0.9774 &0.9774 &0.9774	\\\hline
$|V_3|$&0.9953  &0.9953 &0.9953 &0.9953	\\\hline
$|V_4|$&0.9447  &0.9447 &0.9447 &0.9447 	\\\hline
\end{tabular}
\vspace{1mm}

\caption{$P_6=0.30$ (failure of first-order relaxation)} \label{t2}
\vspace{-1mm}
\centering
\renewcommand{\arraystretch}{1.2}
\begin{tabular} {|c||c|c|c|c|} \hline 
 Voltage        & Pad\'{e}      & Newton & SDP  & SDP \\
 Magnitude      & Approx.      & Raphson & (1st order)  & (2nd order) \\\hline
$|V_1|$&0.9217  &0.9217&-&0.9217	\\\hline
$|V_2|$&0.9640  &0.9640&-&0.9640	\\\hline
$|V_3|$&0.9897  &0.9897&-&0.9897	\\\hline
$|V_4|$&0.9403  &0.9403 &-&0.9403 	\\\hline
\end{tabular}
\vspace{1mm}

\caption{$P_6=0.75$ (false solution of moment relaxation)} \label{t3}
\vspace{-1mm}
\centering
\renewcommand{\arraystretch}{1.2}
\begin{tabular} {|c||c|c|c|c|} \hline 
 Voltage        & Pad\'{e}      & Newton & SDP  & SDP \\
Magnitude      & Approx.      & Raphson & (1st order)  & (2nd order) \\\hline
$|V_1|$&0.7613  &0.7613&-&0.8504	\\\hline
$|V_2|$&0.8658  &0.8658&-&0.9456	\\\hline
$|V_3|$&0.9210  &0.9210&-&0.7960	\\\hline
$|V_4|$&0.8888  &0.8888 &-&0.1321 	\\\hline
\end{tabular}
\vspace{1mm}

\caption{$P_6=1.00$ (non-convergence of Newton-Raphson) } \label{t4}
\vspace{-1mm}
\centering
\renewcommand{\arraystretch}{1.2}
\begin{tabular} {|c||c|c|c|c|} \hline 
 Voltage        & Pad\'{e}      & Newton & SDP  & SDP \\
 Magnitude      & Approx.      & Raphson & (1st order)  & (2nd order) \\\hline
$|V_1|$&0.5657  &-&-&0.8609	\\\hline
$|V_2|$&0.7546  &-&-&0.9405	\\\hline
$|V_3|$&0.8394  &-&-&0.8178	\\\hline
$|V_4|$&0.8319  &-&-&0.1294	\\\hline
\end{tabular}
\vspace{1mm}

\caption{$P_6=1.02$ (false solution of Newton-Raphson) } \label{t5}
\vspace{-1mm}
\centering
\renewcommand{\arraystretch}{1.25}
\begin{tabular} {|c||c|c|c|c|} \hline 
 Voltage        & Pad\'{e}      & Newton & SDP  & SDP \\
 Magnitude      & Approx.      & Raphson & (1st order)  & (2nd order) \\\hline
$|V_1|$&0.5355  &0.1520&-&0.8575	\\\hline
$|V_2|$&0.7380  &0.0673&-&0.9380	\\\hline
$|V_3|$&0.8283  &0.7376&-&0.8167	\\\hline
$|V_4|$&0.8247  &0.7757&-&0.1296	\\\hline
\end{tabular}
\vspace{3mm}

\caption{$P_6=1.12$ (non-existence of a physical solution)} \label{t6}
\vspace{-1mm}
\centering
\renewcommand{\arraystretch}{1.25}
\begin{tabular} {|c||c|c|c|c|} \hline 
 Voltage        & Pad\'{e}      & Newton & SDP  & SDP \\
 Magnitude     & Approx.      & Raphson & (1st order)  & (2nd order) \\\hline
$|V_1|$&-&0.1242&-&0.8363	\\\hline
$|V_2|$&-&0.0680&-&0.9234	\\\hline
$|V_3|$&-&0.7224&-&0.8086	\\\hline
$|V_4|$&-&0.7609&-&0.1308	\\\hline
\end{tabular}
\vspace{-4mm}
\end{table}

\begin{figure*}
\centering
\begin{circuitikz}[scale=1, transform shape]

\path[draw,line width=4pt] (0,0) -- (0,2);
\draw (-1.6,2.2) node[right] {\text{Reference}};
\draw (-1.6,1.8) node[right] {\text{(Slack)}};
\draw (-1.6,-0.3) node[right] {$|V_r| = 1.00$};
\draw (-1.6,0.2) node[right] {$\theta_r = 0^\circ$};
\path[draw,line width=1pt] (-0.5,1) -- (0,1);
\draw[line width=1] (-1,1) circle (0.5);
\draw (-1,1) node{\scalebox{2}{$\sim$}};

\path[draw,line width=4pt] (4,0) -- (4,2);
\draw (3.8,2.25) node[right] {1};
\draw[-latex,line width=1pt] (4,1)-- (5,1);
\draw (4.90,0.98) node[right] {$0.20+j0.10$};

\path[draw,line width=4pt] (9,0) -- (9,2);
\draw (8.8,2.25) node[right] {3};
\draw[-latex,line width=1pt] (9,1)-- (10,1);
\draw (9.9,0.98) node[right] {$0.20+j0.10$};

\path[draw,line width=4pt] (14,0) -- (14,2);
\draw (13.8,2.25) node[right] {6};
\path[draw,line width=1pt] (14.5,1) -- (14,1);
\draw[line width=1] (15,1) circle (0.5);
\draw (15,1) node{\scalebox{2}{$\sim$}};
\draw (14,0.25) node[right] {$|V_6| = 1.10$};
\draw (14,1.75) node[right] {$P_6 = 1.00$};

\path[draw,line width=4pt] (4,-4) -- (4,-2);
\draw (3.8,-1.75) node[right] {2};
\draw[-latex,line width=1pt] (4,-3)-- (5,-3);
\draw (4.9,-3.02) node[right] {$0.10+j0.05$};

\path[draw,line width=4pt] (9,-4) -- (9,-2);
\draw (8.8,-1.75) node[right] {5};
\path[draw,line width=1pt] (9,-3) -- (9.5,-3);
\draw[line width=1] (10,-3) circle (0.5);
\draw (10,-3) node{\scalebox{2}{$\sim$}};
\draw (9,-3.75) node[right] {$|V_5| = 1.10$};
\draw (9,-2.25) node[right] {$P_5 = 1.00$};

\path[draw,line width=4pt] (14,-4) -- (14,-2);
\draw (13.8,-1.75) node[right] {4};
\draw[-latex,line width=1pt] (14,-3)-- (15,-3);
\draw (14.9,-3.02) node[right] {$0.20+j0.10$};
%

\path[draw,line width=1pt] (0,1)-- (4,1);
\draw (2,1.55) node[below] {$0.70+j0.40$};

\path[draw,line width=1pt] (4,1.66)-- (9,1.66);
\draw (6.5,2.20) node[below] {$0.50+j0.50$};

\path[draw,line width=1pt] (9,1.66)-- (14,1.66);
\draw (11.5,2.20) node[below] {$0.40+j0.50$};

\path[draw,line width=1pt] (4,0.33) -- (4.5,0.33);
\path[draw,line width=1pt] (4.5,0.35) -- (4.5,-2.35);
\path[draw,line width=1pt] (4.5,-2.33) -- (4,-2.33);
\draw (4.75,-1) node[rotate=-90] {$0.40+j0.60$};

\path[draw,line width=1pt] (9,0.33) -- (8.5,0.33);
\path[draw,line width=1pt] (8.5,0.35) -- (8.5,-2.35);
\path[draw,line width=1pt] (8.5,-2.33) -- (9,-2.33);
\draw (8.25,-1) node[rotate=90] {$0.30+j0.50$};

\path[draw,line width=1pt] (4,-3.66)-- (9,-3.66);
\draw (6.5,-3.66) node[below] {$0.30+j0.60$};

\path[draw,line width=1pt] (14,0.33) -- (13.5,0.33);
\path[draw,line width=1pt] (13.5,0.35) -- (13.5,-2.35);
\path[draw,line width=1pt] (13.5,-2.33) -- (14,-2.33);
\draw (13.25,-1) node[rotate=90] {$0.60+j0.80$};

\path[draw,line width=1pt] (9,0.334) -- (9.51,0.334);
\path[draw,line width=1pt] (9.5,0.334) -- (13.5,-3.667);
\path[draw,line width=1pt] (13.49,-3.667) -- (14,-3.667);
\draw (11.5,-1.25) node[rotate=-45] {$0.50+j0.80$};

\end{circuitikz}
\caption{7-bus network\vspace{7mm}}
\label{7b}

\centering
        \includegraphics [scale=1.1,trim=0.0cm 0.cm 0.0cm 0.0cm,clip]{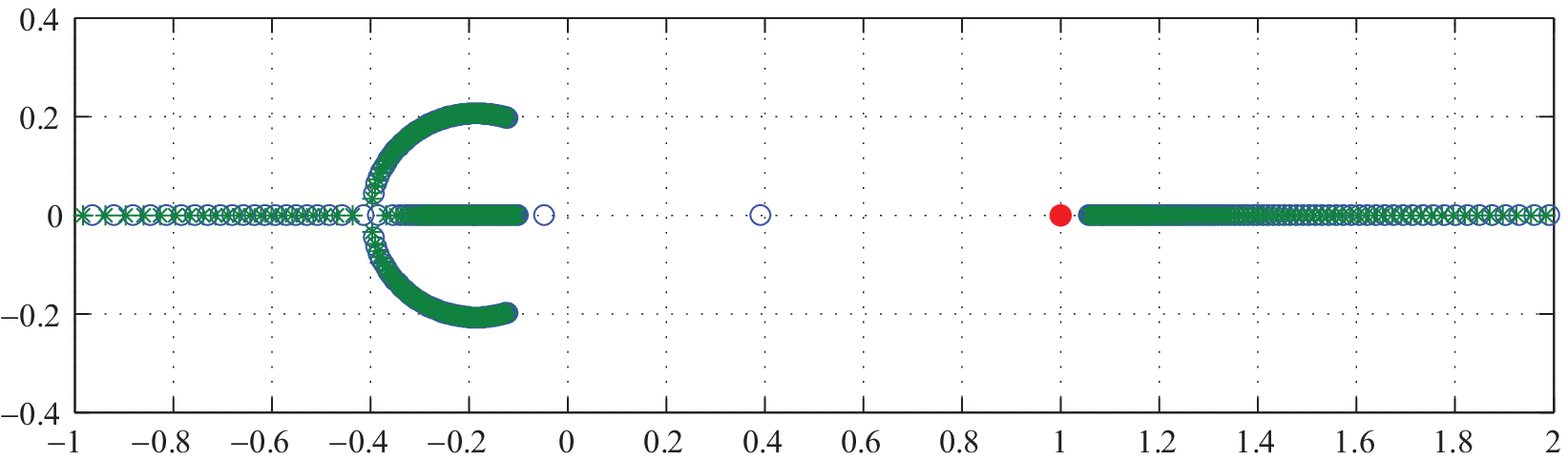}
         \caption{Zero-pole distribution of PA[1000/1000] depicting the analytic structure of voltage phasors in the network of Fig.~\ref{7b} (corresponding to
the embedding approach defined in Section III of Part I of this paper).\vspace{5mm}}\label{PA-100p6}

         \centering
        \includegraphics [scale=1.1,trim=0.0cm 0.cm 0.0cm 0.0cm,clip]{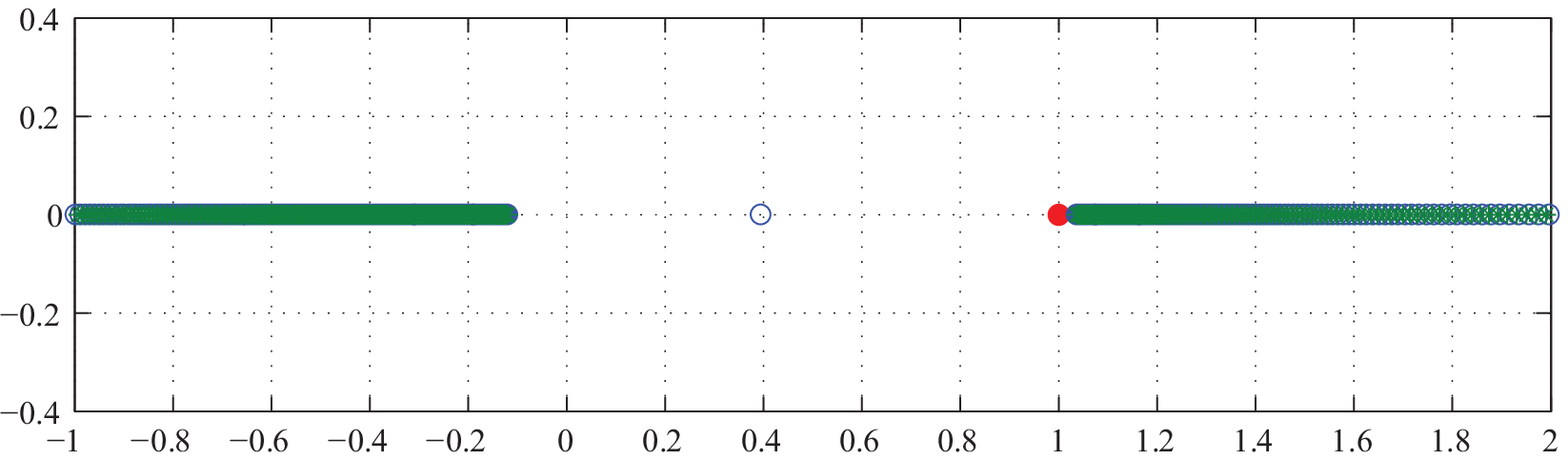}
         \caption{Zero-pole distribution of PA[1000/1000] depicting the analytic structure of voltage phasors in the network of Fig.~\ref{7b} (corresponding to
the embedding approach defined in Section IV of Part I of this paper).\vspace{0mm}}\label{PA-100p6_2nd}
\end{figure*}

\section{Superiority of the embedding approach over Newton-Raphson and Semidefinite Relaxation Methods}

Figure~\ref{7b} shows a 7-bus network with 4 load (PQ) buses labeled 1-4, two generator (PV) buses labeled 5 and 6 and a reference (slack) bus. All values are in per unit. Line and load parameters are indicated as complex quantities. The generator voltage magnitudes are controlled at 1.10 and their active power output is 1.00. Newton-Raphson fails to solve this problem as it does not converge with a flat start, i.e. when initialized with all phase angles set to zero and all PQ voltage magnitudes set to $V_r$. The first-order semidefinite relaxation also fails as it is not tight enough and the second-order (moment-based) relaxation finds a false solution. In contrast the embedding framework, in its both approaches discussed in Sections III and IV of Part I of this paper, finds the true solution and, as the zero-pole distributions in Figures~\ref{PA-100p6} and~\ref{PA-100p6_2nd} clearly demonstrate, the operating point is on the stable branch and still has some margin to the point of voltage collapse.

\begin{figure*}
        \centering
  \subfloat[Second-order semidefinite (moment-based) relaxation solutions (green) versus PA solutions (red)]{\label{P6V3a}\includegraphics [scale=0.88,trim=0.0cm 0.cm 0.0cm 0.0cm,clip]{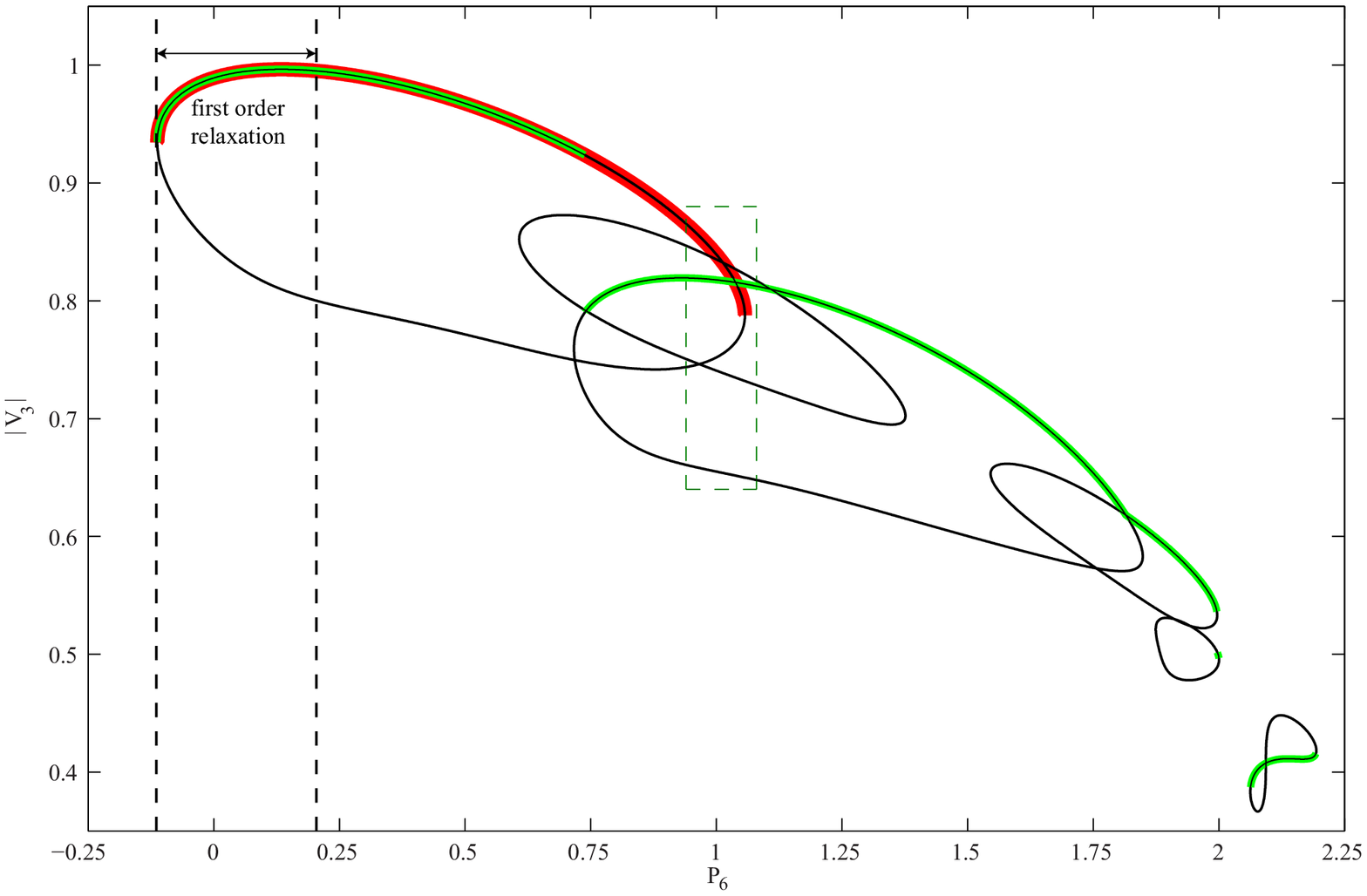}}
                   \\
         \subfloat[Newton-Raphson solutions (blue) versus PA solutions (red)]{\label{P6V3b}\includegraphics  [scale=0.88,trim=0.0cm 0.0cm 0.0cm 0.0cm,clip]{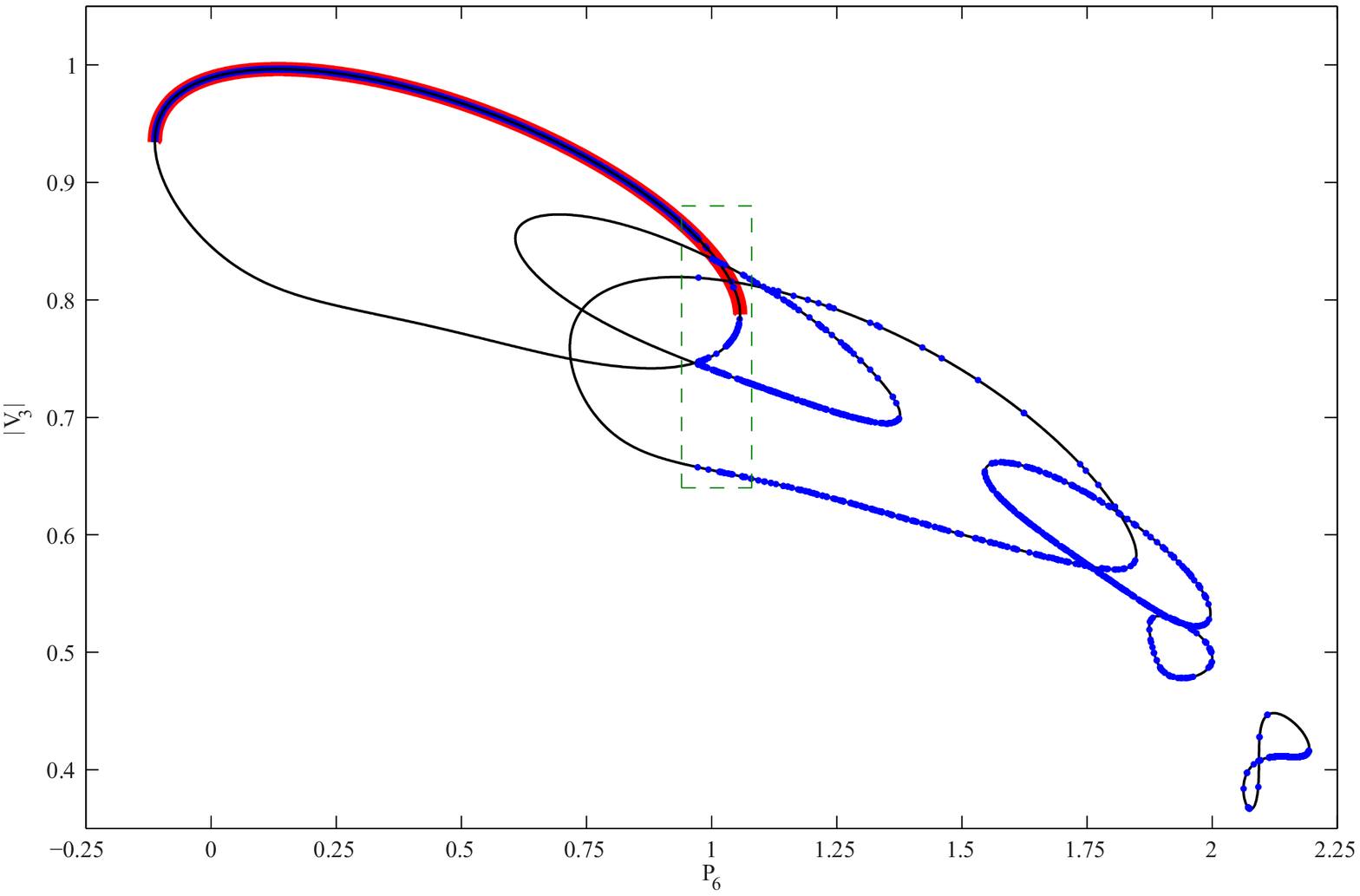}}

         \caption{Superiority of the embedding framework over Newton-Raphson and SDP methods shown in the context of the network of Fig.~\ref{7b}} \label{P6V3}
\end{figure*}

\begin{figure}[h]
        \centering
           \includegraphics  [scale=0.73,trim=0.10cm 0.0cm 0.0cm 0.0cm,clip]{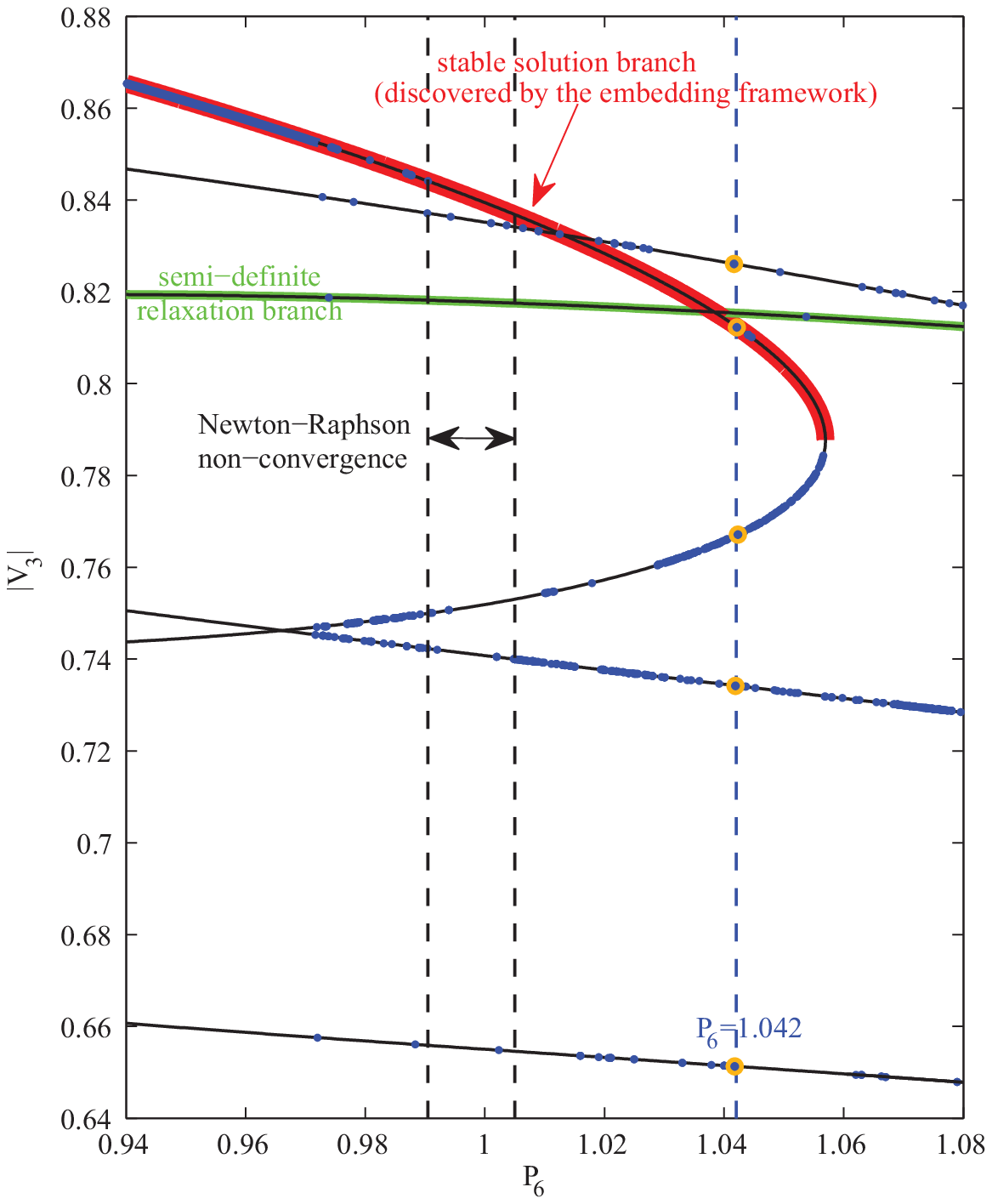}
           \caption{PA solutions (red) versus moment relaxation solutions (green) and Newton-Raphson solutions (blue)\vspace{-2mm}}\label{zoom}
\end{figure}

Now consider the power flow in its polar form. The variables are phase angles of the 6 buses and voltage magnitudes of the 4 load buses. The power flow has 10 equations, relating the active power of the 6 buses and the reactive power of the 4 load buses to phases angles and voltage magnitudes. To better contrast these methods we free a single parameter, $P_6$, the active power generated at bus 6. Each equation defines a hypersurface in $\R^n$ where $n=6+4+1$. The intersection of these hypersurfaces, once projected onto the joint space of the freed parameter and a given variable, yields a series of curves in $\R^2$. Figure~\ref{P6V3} shows these curves in black color in the $(P_6,|V_3|)$ space. The stable operating points of the network of Figure~\ref{7b} can only be realized on the segment that is highlighted in red. This segment is consistently found by the embedding framework for all values of $P_6 \in [-0.114 \quad 1.057]$. However Newton-Raphson and semidefinite relaxation methods concurrently find the stable branch only on a small subset of this interval (Table~\ref{t1}). Beyond $P_6=0.204$ the first-order relaxation fails (Table~\ref{t2}). Beyond $P_6=0.739$ the second-order relaxation finds the false branches (Table~\ref{t3}). These branches are highlighted in green in Figure~\ref{P6V3}a. Newton-Raphson convergence becomes erratic beyond $P_6=0.973$ (Table~\ref{t4}). As Figure~\ref{P6V3}b shows it either does not converge for certain values of $P_6$ (Table~\ref{t4}) or it converges to low-voltage, physically unrealizable and thus false operating points (Table~\ref{t5}). Beyond $P_6=1.057$ there is no physically meaningful solution and the embedding framework returns no solution whereas both Newton-Raphson and moment-based relaxation find false solutions (Table~\ref{t6}). It should also be noted that Newton-Raphson is not robust even in finding low-voltage solutions. This is shown in Table~\ref{t6} and highlighted in Figure~\ref{zoom} where small perturbations at $P_6=1.042$ results in Newton-Raphson finding different branches or not converging at all. For industrial applications power flow parameters and states are typically expressed in 2, 3 and rarely 4 significant digits. Hence the set of values $(1.0416,...,1.0424)$ can be rounded to 4 significant digits and represented as $1.042$ but applying Newton-Raphson to this set yields five topologically distinct solutions as well as non-convergence. Figure~\ref{zoom} shows the detail of the region highlighted by a dashed green box in Figures~\ref{P6V3}a and~\ref{P6V3}b. This region contains the saddle-node bifurcation at $P_6=1.057$ and presents a visual contrast between the performance of the embedding framework and those of Newton-Raphson and moment-based relaxation. Notice that Newton-Raphson finds solutions, mostly false, on all 6 solution branches as shown in Figure~\ref{zoom} whereas moment-based relaxation consistently finds the false branch that Newton-Raphson rarely discovers.

\begin{table*}
\caption{Erratic convergence of Newton-Raphson at $P_6=1.0416...1.0424\approx1.042$} \label{ext}
\centering
\renewcommand{\arraystretch}{1.15}
\begin{tabular} {|c||c|c|c|c|c|c|c|c|c|} \hline
$P_6$  &1.0416&	1.0417&	1.0418&	1.0419&	1.0420&1.0421&	1.0422&	1.0423&	1.0424\\\hline \hline
$|V_1|$&0.3189&	0.4696&	0.2889&	0.1438&	-&0.1438&	0.4887&	0.3210&	0.1436\\\hline
$|V_2|$&0.6256&	0.0520&	0.6478&	0.0675&	-&0.0675&	0.7126&	0.6266&	0.0675\\\hline
$|V_3|$&0.7667&	0.8261&	0.6513&	0.7342&	-&0.7342&	0.8123&	0.7671&	0.7341\\\hline
$|V_4|$&0.7902&	0.8221&	0.1472&	0.7725&	-&0.7725&	0.8147&	0.7904&	0.7724\\\hline
\end{tabular}
\vspace{-5mm}
\end{table*}

\subsection{What about modifications to Newton-Raphson or other heuristic approaches?}

\begin{figure}
        \centering
\includegraphics[scale=0.46,trim=0.65cm 0cm 0.0cm 0.0cm,clip]{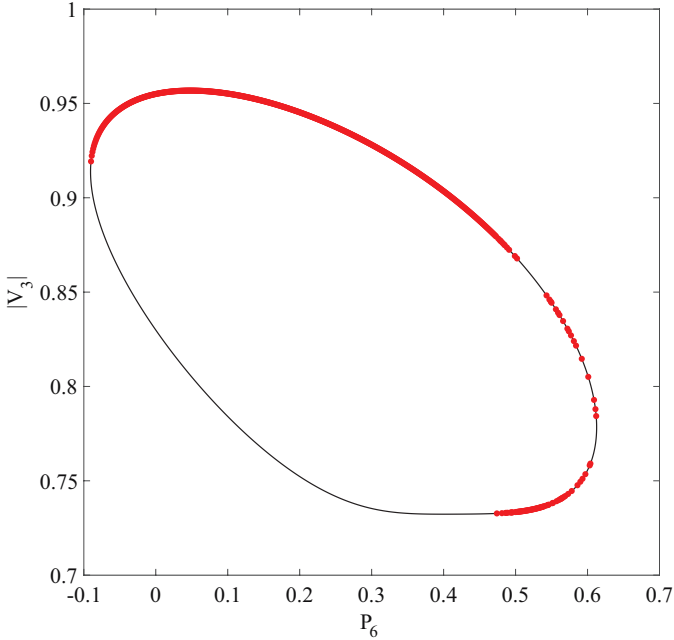}
\caption{Problematic convergence of an alternative Jacobian formulation for polar Newton-Raphson when the line connecting bus 1 and bus 2 in Fig.~\ref{7b} is removed.\vspace{-5mm}}\label{NRm}
\end{figure}

In the previous analysis Newton-Raphson solutions are initialized from flat-start with the Jacobian matrix constructed according to~\eqref{polar}. However there are alternative formulations for the Jacobian matrix that result in drastically different convergence properties. It has been brought to our attention that if the diagonal elements of submatrices $\partial P/\partial |V|$ and $\partial Q/\partial |V|$, i.e.~\eqref{dpdv} and~\eqref{dqdv} are replaced with their equivalent forms $P_i/|V_i|+G_{ii}|V_i|$ and $Q_i/|V_i|-B_{ii}|V_i|$, the Newton-Raphson can obtain the stable branch all the way up to the onset of voltage collapse at $P_6=1.057$. In this particular case, this alteration of the Jacobian matrix partially addresses the convergence issues but it does not resolve the issue of a finding false solution (for example, check the solution at $P_6=1.090$). In other cases, this alteration may result in an even more problematic convergence pattern than the original formulation of the Jacobian. For example consider the 7-bus network of Figure~\ref{7b} with the line connecting bus 1 and bus 2 removed. Figure~\ref{NRm} shows the problematic convergence of this alternative formulation of the Jacobian matrix in vicinity of the saddle-node bifurcation point at $P_6=0.612$. Notice that due to the line removal the range of $P_6$ values with a stable operation has shrunk significantly compared to Figure~\ref{zoom}. The alteration of the Jacobian matrix does not result in any convergence for all values of $P_6 \in [0.502 \quad 0.537]$ whereas with the original Jacobian matrix Newton-Raphson converges to the stable branch for all feasible values of $P_6$. With further modifications to the network of Figure~\ref{7b} one can find numerous cases where alternative forms of Jacobian matrix simultaneously fail to yield the true solution.

\begin{subequations}\label{polar}
\begin{align}
\label{dpda}
&\displaystyle \frac{\partial P_i}{\partial \theta_i} = |V_i| \sum_{k\in \mathcal{N}(i)} \left |V_k|(B_{ik}\cos\theta_{ik}-G_{ik}\sin\theta_{ik} \right)\\
&\displaystyle \frac{\partial P_i}{\partial \theta_k} = |V_i| |V_k|(G_{ik}\sin\theta_{ik}-B_{ik}\cos\theta_{ik})\\
\label{dqda}
&\displaystyle\frac{\partial Q_i}{\partial \theta_i} = |V_i| \sum_{k\in \mathcal{N}(i)} \left |V_k|(G_{ik}\cos\theta_{ik}+B_{ik}\sin\theta_{ik} \right)\\
&\displaystyle \frac{\partial Q_i}{\partial \theta_k} = -|V_i| |V_k|(G_{ik}\cos\theta_{ik}+B_{ik}\sin\theta_{ik} )\\
\label{dpdv}
&\displaystyle \frac{\partial P_i}{\partial |V_i|} = \hspace{-2mm} \sum_{k\in \mathcal{N}(i)} \hspace{-2mm} \left |V_k|(G_{ik}\cos\theta_{ik}+B_{ik}\sin\theta_{ik} \right)+2G_{ii}|V_i|\\
&\displaystyle \frac{\partial P_i}{\partial |V_k|} = |V_i|(G_{ik}\cos\theta_{ik}+B_{ik}\sin\theta_{ik})\\
\label{dqdv}
&\displaystyle \frac{\partial Q_i}{\partial |V_i|} = \hspace{-2mm}\sum_{k\in \mathcal{N}(i)} \hspace{-2mm} \left |V_k|(G_{ik}\sin\theta_{ik}-B_{ik}\cos\theta_{ik} \right)-2B_{ii}|V_i|\\
&\displaystyle \frac{\partial Q_i}{\partial |V_k|} = |V_i|(G_{ik}\sin\theta_{ik}-B_{ik}\cos\theta_{ik})
\end{align}
\end{subequations}

In industrial applications, Newton-Raphson is often applied as the corrector part of a predictor-corrector continuation and there are other heuristic approaches to improve the convergence of this method. For example, the Newton's method with an optimal multiplier successfully solves the power flow for all cases presented in Tables~\ref{t1}-\ref{t5}. Similar to the modification of the Jacobian matrix there is no guarantee that these heuristic approaches always succeed in finding the true solution and avoid convergence to false solutions. Furthermore the heuristic approaches typically increase the computational cost. For example in the context of the 7-bus network of Figure~\ref{7b}, one can solve the power flow for $P_6=0$ and then use the null vector of the augmented Jacobian which is tangent to the black curve at $P_6=0$ for predicting the power flow variables at $P_6=1.00$. This prediction can then be used to initialize the Newton's method to obtain the solution corresponding to $P_6=1.00$. However this heuristic approach finds the false solution\footnote{There is a distinction between {\em warm-starting} the Newton's method and the classical power flow continuation method. The latter can at least theoretically find a stable operating point, given a true starting point, in the limit of infinitesimal step size. This follows from the smoothness of the equations and the implicit function theorem. This method however, suffers from several serious practical limitations especially for large systems and when the power flow parameter space is non-convex.}. For it to be successful the intermediary solution needs to be chosen closer to the target point of $P_6=1.00$, i.e. closer to the point of voltage collapse and it is not known a priori what value of $P_6$ would work. Moreover this approach requires solving the power flow at least twice and possibly even more. These problems highlight the fundamental limitations of solvers that use Newton-Raphson and its variants. Commercial power flow developers claim that a combination of heuristic methods reduces the likelihood of power flow solution failure~\cite{PSSE}. In other words, if the solution does exist and one method fails to obtain it, at least some other established method will most likely succeed. Even if one accepts this claim, the question remains ``which solution?"

%
%
%
\section{Detecting the Closest Saddle-Node and Limit-Induced Bifurcation}

As demonstrated in Part I of this paper, the analytically continued solution at $z=1$ is guaranteed by the Stahl's theory to be on the same algebraic sheet as the one containing the trivial stable solution, i.e. the zero-current solution. This extraordinary strength of complex analysis can be tapped for voltage collapse studies where it is critical to detect the saddle-node or limit-induced bifurcations in the power flow parameter space. Classical homotopy methods based on predictor-corrector algorithms start from a known stable solution to construct a path toward the actual solution, the feasibility of which is not known beforehand. If the solution path passes through a saddle-node bifurcation a single eigenvalue changes sign. Therefore by checking the eigenstructure of the Jacobian matrix along the path, one can detect saddle-node bifurcation and stop the homotopy process. This process has some downsides. First, it may not be easy to find a suitable initial solution. Next, the underlying predictor-corrector algorithms can be very cumbersome. Finally, in the presence of various types of network controllers and their limits, the analysis of the eigenstructure of the Jacobian is often too complicated to be practical for large networks. The situation is no better in optimization-based methods including semidefinite programming where the power flow solution space boundary is relaxed and techniques developed based on real-algebraic geometry are employed to constrain the relaxed boundary.  In contrast to these methods as the power flow parameters are perturbed the embedding framework always finds the stable solution and detects the saddle-node bifurcation. This coincides with a non-trivial monodromy in the complex plane as a result of the branch points of analytic functions reaching $z=1$. The detection of saddle-node bifurcation is done either numerically through inspecting the PA solution as the order of approximants is increased or by inspecting the zero-pole distribution of the approximant. Table~\ref{tpade} shows the PA solution for the network of Figure~\ref{7b} for an increasing order of diagonal Pad\'{e} approximants. This is based on the embedding approach defined in Section III of Part I of this paper. Table~\ref{tpade_2nd} shows the same process based on the second embedding approach defined in Section IV of Part I of this paper. It is clear that in both approaches, for PA[50/50] and higher orders, the accuracy of the computed voltage magnitudes does not change within 4 digits past the decimal point{\footnote{We have noticed that for some larger networks the first approach requires a much higher order of diagonal Pad\'{e} approximants to yield similar results in comparison with the second approach. Hence, the second approach might be more advantageous from a computational point of view. We will discuss this in Part III of the paper. Nonetheless, both approaches have the same capabilities when it comes to detecting the onset of voltage collapse and providing a reliable stability margin.}. However if the power flow is infeasible as the order of the diagonal Pad\'{e} approximant increases the approximated values no longer converge. Similar conclusion can be made based on the location of the closest branch point on the positive real axis. If this branch point moves closer to the origin past the point $z=1$ then the power flow is infeasible. This aspect of the embedding framework is particularly promising for voltage collapse studies since for heavily-loaded operating points, the location of the critical branch point can be determined fairly accurately at low orders of Pad\'{e} approximants. For example in Figure~\ref{PA-100p6} corresponding to the 7-bus network of Figure~\ref{7b} PA[50/50] is sufficient to determine with 4 digits of accuracy the location of this branch point on the positive real axis. Hence the embedding framework not only solves or detects infeasibility of the power flow but also offers, as a byproduct, an efficient and fast proximity index to voltage collapse. We should emphasize that zero-pole inspection is the superior method to detect saddle-node bifurcation as opposed to numerical inspection of the solution for an increasing order of Pad\'{e} approximants. The reason is that as the branch point approaches $z=1$, the density of the zeros and poles of the approximant at this branch point increases. Therefore in stressed networks the zero-pole distribution of diagonal Pad\'{e} approximants of small orders can accurately pinpoint the location of the critical branch point.


\begin{table}[h]
\caption{PA solution of the network of Fig.~\ref{7b} for an increasing order of diagonal Pad\'{e} approximants
(corresponding to the embedding approach defined in Section III of Part I of this paper).} \label{tpade}
\centering
\renewcommand{\arraystretch}{1.20}
\begin{tabular} {|c||c|c|c|c|c|} \hline 
 Voltage        & PA      & PA& PA  & PA & PA \\
 Magnitude     & [20/20]      & [30/30] & [40/40]  & [50/50] & [60/60] \\\hline
$|V_1|$&0.5703&0.5664&0.5658&0.5657&0.5657	\\\hline
$|V_2|$&0.7572&0.7549&0.7546&0.7546&0.7546	\\\hline
$|V_3|$&0.8408&0.8396&0.8394&0.8394&0.8394	\\\hline
$|V_4|$&0.8328&0.8321&0.8319&0.8319&0.8319	\\\hline
$|V_5|$&1.1001&1.0999&1.1000&1.1000&1.1000	\\\hline
$|V_6|$&1.0997&1.1000&1.1000&1.1000&1.1000	\\\hline
\end{tabular}
\vspace{5mm}

\caption{PA solution of the network of Fig.~\ref{7b} for an increasing order of diagonal Pad\'{e} approximants
(corresponding to the embedding approach defined in Section IV of Part I of this paper).} \label{tpade_2nd}
\centering
\renewcommand{\arraystretch}{1.20}
\begin{tabular} {|c||c|c|c|c|c|} \hline 
 Voltage        & PA      & PA& PA  & PA & PA \\
 Magnitude     & [20/20]      & [30/30] & [40/40]  & [50/50] & [60/60] \\\hline
$|V_1|$&0.5678&0.5659&0.5658&0.5657&0.5657	\\\hline
$|V_2|$&0.7557&0.7547&0.7546&0.7546&0.7546	\\\hline
$|V_3|$&0.8401&0.8395&0.8394&0.8394&0.8394	\\\hline
$|V_4|$&0.8323&0.8319&0.8319&0.8319&0.8319	\\\hline
$|V_5|$&1.0999&1.1000&1.1000&1.1000&1.1000	\\\hline
$|V_6|$&1.0999&1.1000&1.1000&1,1000&1.1000	\\\hline
\end{tabular}
\end{table}

\begin{figure}
        \centering
        \includegraphics [scale=0.76,trim=1.7cm 0.cm 0.0cm 0.0cm,clip]{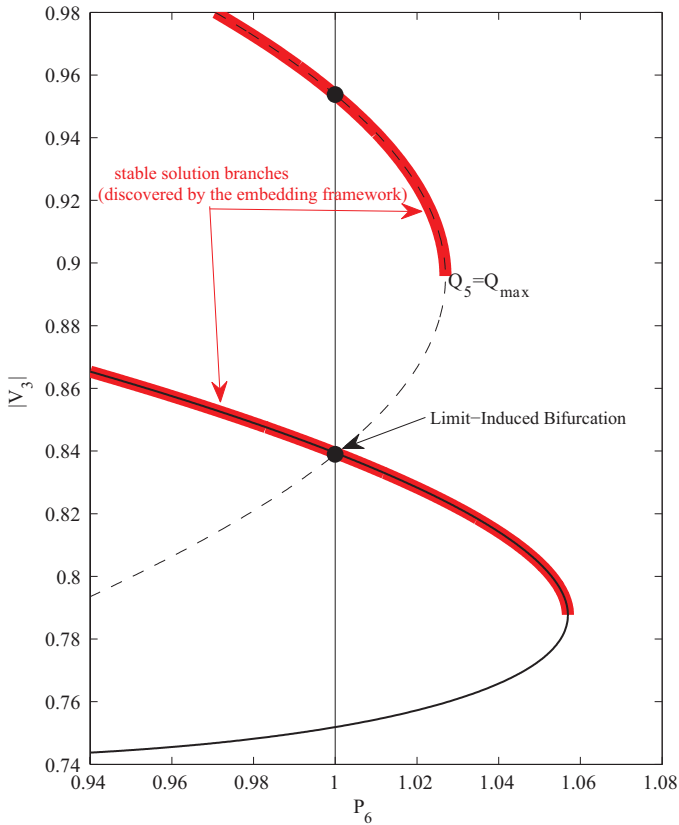}
         \caption{ The embedding framework detecting a limit-induced bifurcation corresponding to the network of Fig.~\ref{7b}\vspace{-7mm}}\label{rlimit}
\end{figure}

\subsection{Limit-induced Bifurcation}

The case of limit-induced bifurcation requires special attention as it may not follow the patterns of saddle-node bifurcation discussed earlier. In a highly-stressed network it is quite likely that as the limit of a controller is enforced the operating point finds itself on the unstable solution branch as shown in Figure~\ref{rlimit}. In such cases the embedding framework finds the limit-enforced solution but on the stable branch. Thus the PA solutions just prior and immediately following the enforcement of the reactive limit are distinctly different. This signifies nothing but the limit-induced bifurcation. In the 7-bus network of Figure~\ref{7b}, the reactive output of generator 5 is $Q_5=0.7466$. Suppose the reactive load at bus 2 is slightly perturbed from $0.0500$ to $0.0518$. This pushes the operating point exactly on the capacitive limit of the generator at $Q_{\text{max}}=0.7500$. Table~\ref{LIBT} lists the PA solutions for the two cases where bus 5 is considered as PV and as PQ. As also highlighted in Figure~\ref{rlimit} by black dots, these two solutions are different and this can only occur when the network experiences a limit-induced bifurcation. Note that the first solution (lower dot), listed under bus 5 as PV, also satisfies the power flow equations under bus 5 as PQ but only as an unstable operating point. However the second solution (upper dot) is also unstable. It would have been stable and thus valid if only bus 5 had been a PQ bus with $P=1.00$ and $Q=0.75$ from the outset. In that case bus 5 voltage would have been 1.2284. However we know that bus 5 is originally a PV bus that is switched to a PQ bus and the fact that $Q=0.75$ is due to PV bus reaching its capacitive limit. The second solution in Table~\ref{LIBT} (upper dot) is unstable for the following reason. The sensitivity of $V_5$ to $Q_5$ is positive here. Thus if somehow the operating point reaches the second solution then $V_5$ start decreasing from 1.2284 to its setpoint at 1.10 with $Q_5$ becoming less capacitive, i.e. coming off the limit. Thus this operating point can occur only in a transient state. One can also check that as $V_5$ starts decreasing toward its setpoint it reaches saddle-node bifurcation. In other words, if we consider bus 5 as PQ and start decreasing $Q_5$ incrementally, the voltage of bus 5 decreases toward 1.10 but at some point, in between, the operating point reaches saddle-node bifurcation. Thus when bus 5 is on its limit neither of the two solutions in Table~\ref{LIBT} are valid. Also note that in Figure~\ref{rlimit} if $P_6>1$ then the embedding framework declares non-existence of the stable solution because the limited solution is unstable for the same reason stated earlier, i.e.  bus 5 is on-limit and has a $\partial{|V|}/\partial{Q}>0$  and $|V|>V_{\text{sp}}$ where $V_{\text{sp}}$ is the setpoint value. In this case by enforcing the limit we obtain an unstable solution. However in general this does not have to be the case. There are two other possibilities.  (1) There may not exist any second solution, stable or unstable, corresponding to a PV bus switched to a PQ bus (e.g. $P_6=1.04$ in Figure~\ref{rlimit}). (2) There may exist another stable solution and this can be verified by checking the sensitivity of the $|V|$ at the now on-limit bus to its reactive power output or tap position that controls that voltage and whether $|V|>V_{\text{sp}}$ or not.

In the second case, the discovery of another stable solution means that this is a perfectly valid operating point that can be reached via a different, often non-trivial, path in the parameter space. Nonetheless the conclusion that limit-induced bifurcation is encountered still holds, i.e. the current path in the parameter space leads to limit-induced bifurcation even though there exists a different path to avoid this phenomenon. \vspace{0mm}

\begin{table}
\caption{PA solutions corresponding to limit-induced bifurcation of the network of Fig.~\ref{7b}.} \label{LIBT}
\centering
\renewcommand{\arraystretch}{1.15}
\begin{tabular} {|c||c|c|} \hline 
 Voltage        & bus 5 as PV     & bus 5 as PQ  \\
 Magnitude     & off-limit     & on-limit \\\hline
$|V_1|$&0.5641&0.7988\\\hline
$|V_2|$&0.7530&0.9625\\\hline
$|V_3|$&0.8390&0.9537\\\hline
$|V_4|$&0.8316&0.8925\\\hline
$|V_5|$&1.1000&1.2284\\\hline
$|V_6|$&1.1000&1.1000\\\hline
\end{tabular}
\vspace{-3mm}
\end{table}

%


\bibliography{bibfile}
\bibliographystyle{ieeetr}


\end{document}